\documentclass[aps, prb, twocolumn,showkeys,showpacs,amsmath,amssymb,superscriptaddress]{revtex4-1}

\usepackage[T1]{fontenc}
\usepackage[latin1]{inputenc}

\usepackage[ngerman,british]{babel}
	\addto\captionsngerman{}

\usepackage{amsmath,amssymb,amsthm,dsfont,mathtools,mathrsfs,bbm}

\usepackage{mdwlist}
\usepackage{paralist}
\usepackage{array}
\usepackage{booktabs}						
\usepackage{dcolumn} 						

\usepackage{graphicx}
\usepackage{wrapfig}

\usepackage{url}

\usepackage{color}
	\definecolor{myblue}{rgb}{0,0.3,0.8}
	\definecolor{mygreen}{rgb}{0,0.5,0}
	\definecolor{myblue}{rgb}{0,0.3,0.8}

\usepackage[version=3]{mhchem}

\usepackage{savesym}

\savesymbol{a}
\savesymbol{b}
\savesymbol{d}
\savesymbol{e}
\savesymbol{k}
\savesymbol{l}
\savesymbol{H}
\savesymbol{L}
\savesymbol{p}
\savesymbol{r}
\savesymbol{s}
\savesymbol{S}
\savesymbol{F}
\savesymbol{t}
\savesymbol{w}
\savesymbol{th}
\savesymbol{div}
\savesymbol{tr}
\savesymbol{Re}
\savesymbol{Im}
\savesymbol{AA}
\savesymbol{SS}
\savesymbol{Si}
\savesymbol{gg}
\savesymbol{ss}

\newcommand{\a}{\alpha}
\newcommand{\b}{\beta}
\newcommand{\e}{\varepsilon}
\newcommand{\d}{\delta}
\newcommand{\del}{\partial}

\newcommand{\g}{\gamma}
\newcommand{\G}{\Gamma}

\newcommand{\r}{\rho}
\newcommand{\s}{\sigma}

\newcommand{\w}{\omega}

\newcommand{\cov}{{\nabla}}

\newcommand{\XX}{{\boldsymbol X}}

\newcommand{\R}{\mathbbm{R}}
\newcommand{\C}{\mathbbm{C}}

\DeclareMathOperator{\Im}{Im}

\newcommand{\T}{\textstyle}

\newlength{\tmplen}

\newcommand{\norm}[1]{\left\lVert#1\right\rVert^2}

\usepackage{bbold}

\begin{document}

\title[Confining Dirac particles]{Confining massless Dirac particles in two-dimensional curved space}

\author{Kyriakos Flouris}\affiliation{ %
ETH
  Z\"urich, Computational Physics for Engineering Materials, Institute
 for Building Materials, Wolfgang-Pauli-Str. 27, HIT, CH-8093 Z\"urich,
 Switzerland}%
\author{Miller Mendoza Jimenez}\affiliation{ %
ETH
  Z\"urich, Computational Physics for Engineering Materials, Institute
 for Building Materials, Wolfgang-Pauli-Str. 27, HIT, CH-8093 Z\"urich,
 Switzerland}%
\author{Jens-Daniel Debus}\affiliation{ %
ETH
  Z\"urich, Computational Physics for Engineering Materials, Institute
 for Building Materials, Wolfgang-Pauli-Str. 27, HIT, CH-8093 Z\"urich,
 Switzerland}%
\author{Hans J. Herrmann}
\affiliation{ %
ETH
  Z\"urich, Computational Physics for Engineering Materials, Institute
 for Building Materials, Wolfgang-Pauli-Str. 27, HIT, CH-8093 Z\"urich,
 Switzerland}%
 \affiliation{ %
Departamento de F\' isica, Universidade do Cear\' a, 60451-970 Fortaleza, Brazil}%
  \affiliation{ %
on leave from C.N.R.S., UMR 7636, PMMH, ESPCI, 10 rue Vauquelin, 75231 Paris Cedex 05, France }%

\begin{abstract}%
Dirac particles have been notoriously difficult to confine.  Implementing a curved space Dirac equation solver based on the quantum Lattice Boltzmann method, we show that curvature in a 2-D space can confine 
a portion of a charged, mass-less Dirac fermion wave-packet. This is equivalent to a finite probability of confining the Dirac fermion within a curved space region. We propose a general power law expression for the probability of confinement with respect to average spatial curvature for the studied geometry.

\end{abstract}

\maketitle

\section{Introduction}

The equations of motion of quasi-relativistic electrons in materials such as carbon nanotubes \cite{nanotubes}, graphene \cite{graphenerev2,miller1,miller2,oliver1,oliver2,illario}, 3D Weyl semimetals \cite{weyl_semimetals, weyl_semimetals_2}  and topological insulators \cite{topo_insulators_1, topo_insulators_2,topo_inslulator_oded}  can be mapped to the equations of motion of relativistic Dirac fermions when considering specific local regions of the Brillouin zone in a periodic lattice \cite{dirac_materials}. These Dirac materials  have revived experimental and theoretical research of quasi-relativistic particles in systems of different dimensions \cite{Dirac_materials_2d, Dirac_material_3d}. 2-dimensional systems are especially of interest due to graphene, 2D topological insulators and the fractional quantum Hall effect \cite{fqhe}. Research is focused on bound states of Dirac particles for example in the context of quantum computing and waveguides \cite{dirac_boundstates_1, dirac_boundstates_2}. 

Mass-less Dirac particles are notoriously difficult to confine.  Studies about transmission of Dirac particles through one-dimensional potentials have shown guided modes using abrupt potentials \cite{waveguide1D,abrupt_potential_2}, however these potentials are not experimentally feasible. In addition, some analytical investigations have indicated confinement only when considering a rotating frame \cite{rot_frame1,rot_frame2}.
Likewise, zero energy or Majorana bound modes have been theoretically predicted for an integrable graphene quantum dot \cite{Bardarson2009} and by forming bielectron vortices\cite{Downing2017}. 

An alternative approach is based on Fermi velocity engineering, for example by embedding Dirac materials on substrates, where the bound states are achieved for specific Fermi velocity geometries \cite{Fermi_vel_eng1, Fermi_vel_eng2}   Furthermore, in the context of quantum dots,  some experimental success has been reported of soft confinement  only by slightly opening the Dirac cone \cite{soft_confinement1,soft_confinement2} and thus deviating from the truly relativistic dispersion.

We propose an alternative geometric approach to confine Dirac particles, namely through static spatial curvature.  Historically there have been other examples of quantum field theories solved in curved space-time in attempts to incorporate some effects of gravity to special-relativistic quantum theories \cite{hawking1975} and these have also been applied to graphene sheets \cite{graphene_hawking}. That is, following Einsteins principle of equivalence, which requires the laws of physics to be the same in all inertial frames, the force of gravity can be modeled as a curvature of space-time. For example, in three dimensions bound states of moving Dirac particles in gravitational fields have been postulated \cite{dirac_boundstates_grav_fields}. Additionally, other effects of curvature, such as energy dissipation \cite{curvature_dissipation}, have also been proposed for classical fluids. Specifically Dirac particles in curved space have been studied in the context of surface electronic structure of topological insulators \cite{Lee2009}, electronic properties of curved graphene sheets with distortion and defects \cite{Cortijo2007}, analytic work of cold Dirac fermions on 2+1D Rindler metric \cite{Boada2011}  and massless Dirac fermions in curved space-time were implemented for understanding the behavior of quantum walks and their use in quantum algorithms \cite{molfetta2013}. In this work we give evidence for spatial confinement of massless Dirac fermions in general 2-dimensional curved-space and on deformed pure mono-layer graphene. 

To this end the transmittance of Dirac fermions is explored numerically using a solver of the Dirac equation developed by Debus et al \cite{JD_thesis}. The method is based on the conceptual similarities between the Dirac and the Boltzmann equations and is an extension of the quantum lattice Boltzmann  method  \cite{succi_qlbm} to curved space.

Firstly we introduce the Dirac equation and its extension to curved space and specifically deformed graphene. In the next section the simulations and results are presented. The paper finishes with a summary and outlook section. An appendix can be found at the end including a more detailed description of the numerical model.

\section{Dirac models in curved space}

The original Dirac equation \cite{dirac} for massless fermions can be written in compact notation:

\begin{equation}
\label{eq:Dirac}
(i \g^{\mu}\del_{\mu})\Psi=0,
\end{equation}
 in natural units such that $\hbar=c=1$ for $\hbar$, Planck's constant and $c$, the speed of light, where $\mu={0,1,2}$ for 2D space-time. $\Psi = (\Psi_a^+, \Psi_a^-) = (\psi_1^+,\psi_2^+,\psi_1^-,\psi_2^-) \in \C^4$ denotes the Dirac spinor for particle, anti-particle $+,-$, and $\g^\mu = \g^\a e_\a^{\ \mu}$ are the generalized space dependent  $\g$-matrices, where $\g^\a \in \C^{4\times 4}$ are the standard Dirac matrices, $e_\a^{\ \mu} $ is the tetrad, which relates the flat Minkowski to the curved space-time with the first and second indexes respectively. 
 
 Here the tetrad is defined by 
\begin{equation*}
e_{\alpha}^{\mu} g_{\mu \nu}e_{\beta}^{\nu}=\eta_{\alpha \beta},
\end{equation*}
where $g_{\mu \nu}$ denotes the metric tensor and $\eta_{\alpha \beta}$ is the 
Minkowski metric. In 2D the tetrad can be computed directly as the square root of the metric. Now defining the covariant derivative as $D_{\mu} \Psi=\del_{\mu} \Psi + \G_{\mu} \Psi$, where $\G_{\mu}$ denotes the spin connection matrices given by 
\begin{align*}
	\Gamma_\mu = - \frac{i}{4} \w_\mu^{\a\b} \s_{\a\b},
\end{align*}
where $\s_{\a\b} = \frac{i}{2} [\g_\a,\g_\b]$ and the spin connection
\begin{align*}
\w_\mu^{\a\b}=\quad&e_\nu^\a \cov_\mu e^{\nu \b}.
\end{align*}
 Using these definitions the Dirac Eq.~(\ref{eq:Dirac}) can be naturally extended to curved space described by a metric tensor $g_{\mu\nu}$ with a covariant derivative as 
\begin{equation}
\label{eq:Dirac}
(i \g^{\mu}D_{\mu})\Psi=0,
\end{equation}
where $D_{\mu}$ is the covariant spinor derivative defined above. The Dirac equation in curved space describes relativistic Dirac particles (e.g. electrons) moving on arbitrary manifold trajectories, Appendix~\ref{app:riemannian}. The  Dirac current is defined by $J^\mu = \overline \Psi \g^\mu \Psi$.

In (2+1) dimensions, of interest in this work, the Dirac matrices can be represented in terms of the standard Pauli matrices $\s^i \in \C^{2\times 2}$, as $\gamma^0=i\s^3, ~ \gamma^1=\s^2 ~ \mathrm{and} ~ \gamma^2=-\s^1$, where $\{\s^i,\s^j\}=\delta^{ij}$. Furthermore, the massless Dirac or Weyl, equation can be expressed in the chiral representation with  $\Psi=(\Psi_L,\Psi_R)$, where  $\Psi_{L/R}=1/2(1 \pm \gamma^5)\Psi$ are decoupled Pauli spinors. In the context of graphene, the general Dirac spinor $\Psi=(\Psi_a^K,\Psi_a^{K'})=(\psi_A^K,\psi_B^K,\psi_A^{K'},\psi_B^{K'})$, for sub-lattices $A,B$ and valleys $K,K'$. Equivalently to the Weyl representation, the two valleys are decoupled from each other, therefore  the spinor can be simplified to the sub-lattice basis, $\Psi=(\psi_A^K,\psi_B^K)$ without loss of generality.

To model the single layer carbon atom honeycomb lattice structure we start from the tight binding Hamiltonian which is constructed assuming superposition of local waves for isolated atoms on a honeycomb lattice\cite{graphene_tb1}. In the low energy limit it has been shown that the tight binding Hamiltonian converges to the Dirac Hamiltonian in the continuum limit,
\begin{equation}
\label{eq:dirac_hamiltonian}
H_D=-i v_f \int \Psi^\dagger \s^i \del_i \Psi d^2x, 
\end{equation}
in natural units, where we have replaced $\gamma^0 \gamma^i$ with $\s^i\in \C^{2\times 2}$, $\Psi$ is in the sub-lattice basis and $v_f$ is the Fermi velocity. The convergence to Eq.~(\ref{eq:dirac_hamiltonian}) can be seen as the Dirac cones in graphene with linear dispersion relation at the conduction and valence band connecting point $E=p$ for $E$, energy and $p$, momentum.

The equation of motion from this Hamiltonian is simply the  Dirac equation. 
In this work, we consider a static space-time metric with trivial time components,
\begin{align*}
	g_{\mu\nu} = 
	\begin{pmatrix}
		1 & 0 \\
		0 & -g_{ij}	
	\end{pmatrix},
\end{align*}
where the latin indices run accros the spatial dimensions. This simplifies the Dirac equation Eq.~(\ref{eq:Dirac}) in (2+1) dimensions to
\begin{equation}
\del_t \Psi + \s^a e_a^i(\del_i + \G_i)\Psi = 0, 
\end{equation}
After addition of external vector and scalar potentials $A_i(x)$ and $V(x)$ respectively as explained in Ref.~\cite{jd_paper}, the Dirac equation takes the following form:
\begin{equation}
\del_t \Psi + \s^a e_a^i(\del_i + \G_i- i A_i)\Psi =\s^3 V \Psi. 
\end{equation}
 For the given metric, the conservation law for the Dirac current can be written as $\del_t \rho + \nabla_i J^i = 0$, where $\r = \Psi^\dagger \Psi \in \R$.

For graphene the effective Dirac Hamiltonian looks like \cite{OLIVALEYVA}: 
\begin{equation}
\label{eq:graphene_hamiltonian}
H^*_D=-i v_f \int \Psi^\dagger \sigma^a (v_a^{*i} \del_i + \G_a^*-i A^*_a)\Psi d^2x,
\end{equation}
where $v_a^{* i}=\d_{a i} + u_{a i} -\beta \e_{a i}$ is the space depended Fermi velocity, $\G^*_a=\frac{1}{2 v_f}\del_j v_a^{* j}$ is a complex vector field and $A^*_a$ is a strain-induced pseudo-vector potential given by $A^*_a=(A_x^*,A_y^*)=\frac{\beta}{2a}(\e_{xx}-\e_{yy},-2\e_{xy}$), $\beta$ is the electron Grueneisen parameter, $a$ the lattice spacing and $\e_{i\jmath}= u_{i\jmath} +\frac{1}{2}\del_i h \del_j h$ the general strain tensor with in-plane, $u_{i\jmath}$ and out of plane, $h$ deformations. The term $u_{a i}$ in $v_a^{* i}$ can be interpreted as the lattice deformation potential term and is purely a geometric consequence. Comparing this to the standard Dirac Hamiltonian in curved space
\begin{equation}
\label{eq:hamiltoniangraphene}
H_D=-i \int \Psi^\dagger \sigma^a e_a^{i}( \del_i + \G_i-i A_i)\Psi \sqrt{g} d^2x,
\end{equation}
we can match both Hamiltonians $H_D$ and $H^*_D$ by fulfilling the following relations:
\begin{equation}
v_a^{*i}= v_f \sqrt{g}e_a^{i}, \ \ \G_a^*= v_f \sqrt{g}e_a^{i}\G_i, \ \ A^*_a= v_f \sqrt{g}e_a^{i}A_i.
\end{equation}
All three can be simultaneously fulfilled by an effective metric tensor derived from the explicit expression of the tetrad \cite{jd_paper}. The effective Dirac model for non-uniformed strained graphene as explained in Ref~\cite{OLIVALEYVA} does not distinguish between the graphene valleys K and K' as it relies on the basic principle that \textit{the theory for graphene under nonuniform strain should describe the particular case of a unifrom strain}, where both Dirac points in the  Brillouin zone are affected the same way and thus considered equivalent. The numerical model implemented is explained in Appendix~\ref{app:method}. 

\section{Simulations and results}

\begin{figure}
\includegraphics[width=\columnwidth, height=\columnwidth]{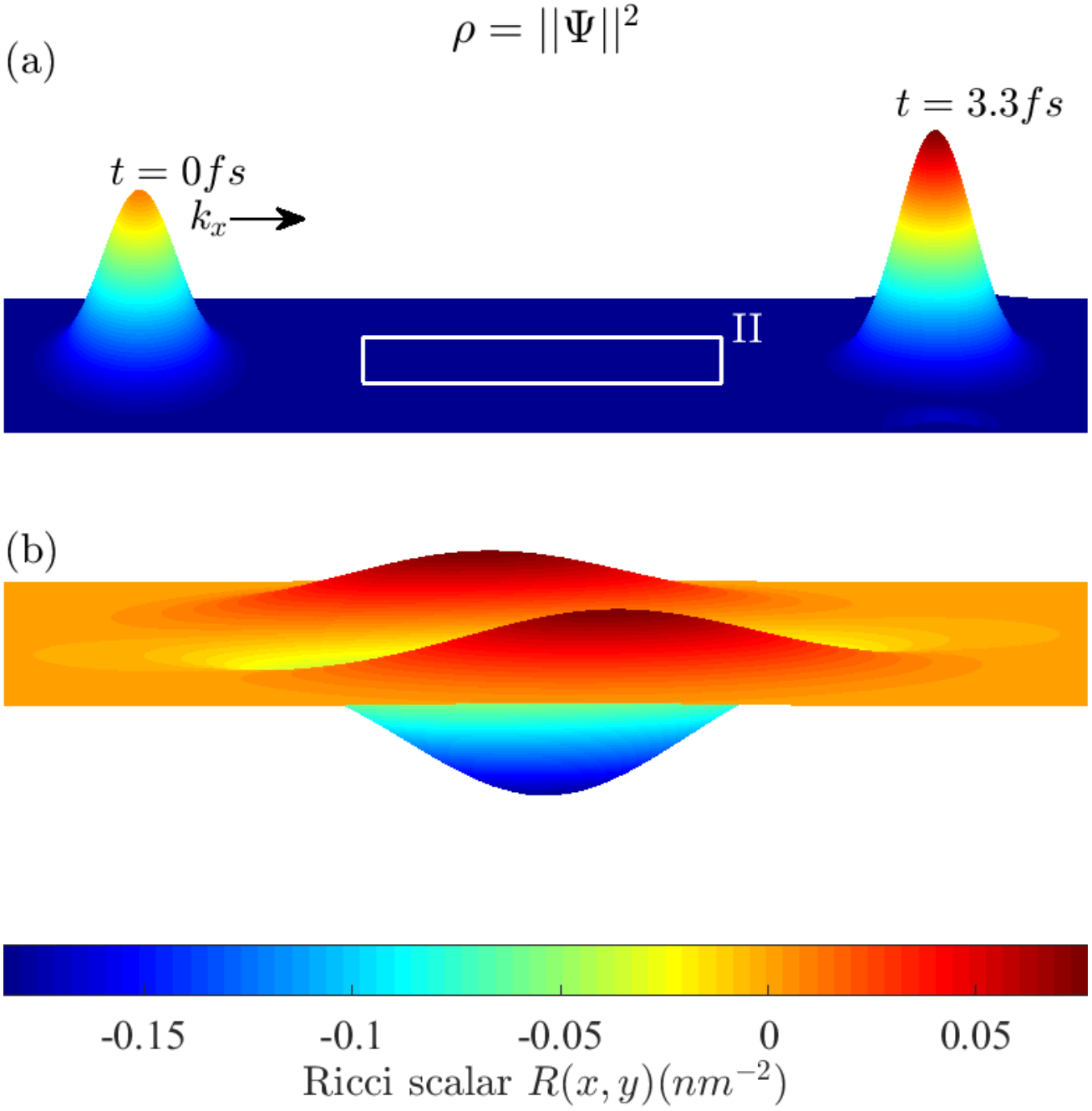}
\caption{\label{fig:schematic} \textbf{(a)} Density of the wave-packet at an initial and a later time-step when the wave-packet exits the curved region in a squeezed Gaussian state  with a higher probability in the center. The white rectangle represents the region $II$ within the numerical domain, where we measure confinement. The arrow denotes the propagation direction. \textbf{(b)} Ricci scalar for $\langle R \rangle= 3.9 \times 10^{-3}nm^{-2}$. }
\end{figure}

\begin{figure}
\includegraphics[width=\columnwidth, height=\columnwidth]{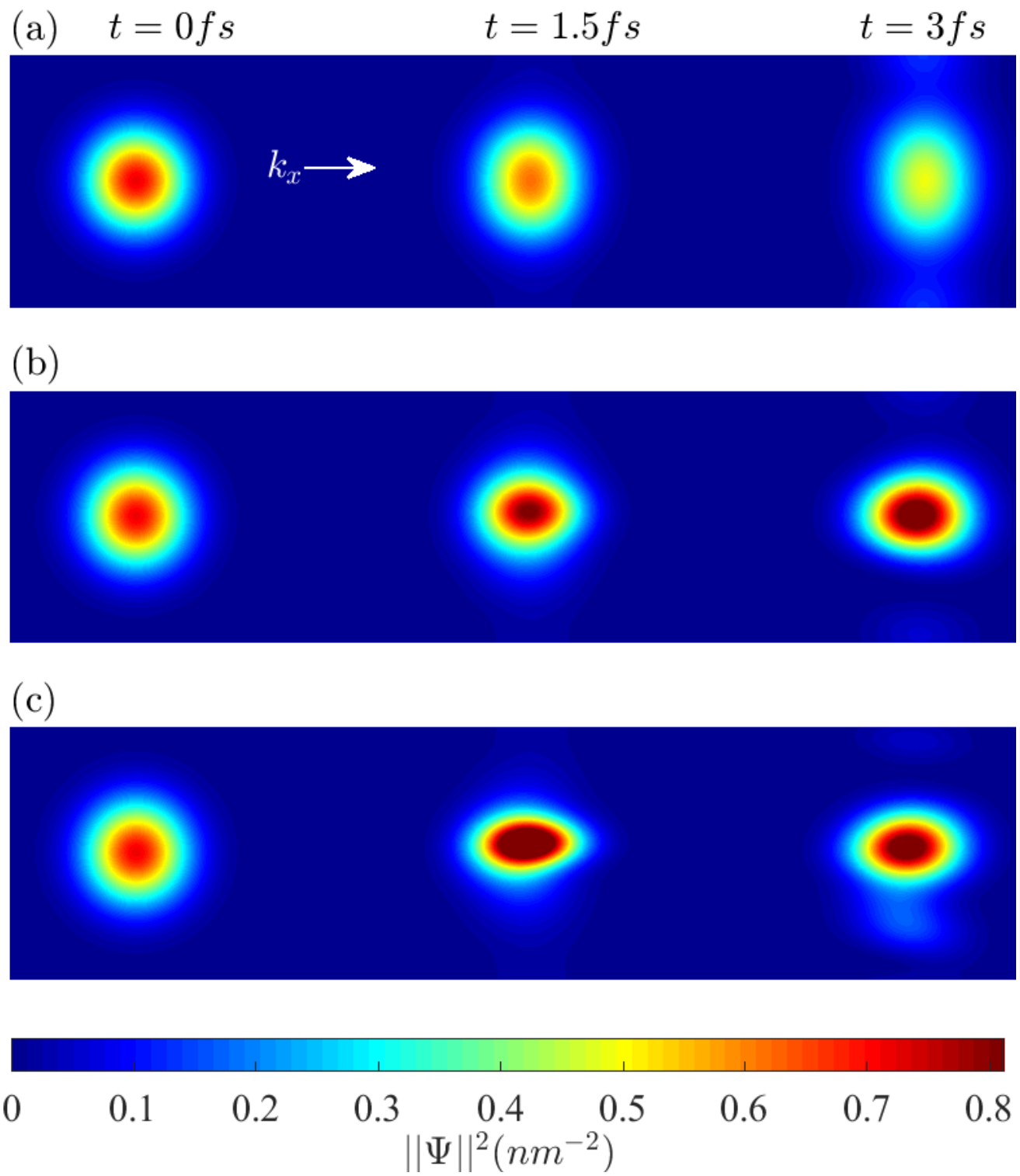}
\caption{\label{fig:squezing} Density plots of the wave-packet at three different time-steps  for $\langle R \rangle= 0, ~ \langle R \rangle= 0.16 \times 10^{-3}nm^{-2}, ~ \langle R \rangle= 3.9 \times 10^{-3}nm^{-2} $ in \textbf{(a), (b)} and  \textbf{(c)} respectively. The arrow denotes the propagation direction. The wave-packet exits the curved space region in a squeezed state.}
\end{figure}

The transmittance of a traveling wave-packet through a curved space obstacle is investigated similarly to Ref.~\cite{miller_klein} for the graphene effective Hamiltonian Eq.~(\ref{eq:graphene_hamiltonian}). A Gaussian wave-packet is initialized as
  \begin{align}
    \Psi(\mathbf{r},\mathbf{k})= \frac{1}{\sqrt{2 \pi \mathcal{\s}^2}} & \begin{pmatrix}
           1 \\
           \lambda e^{i\theta}
         \end{pmatrix}
         e^{-\frac{|\mathbf{r}|^2}{4\mathcal{\s}^2}+i\mathbf{k}\cdot\mathbf{r}},
  \end{align}
where $\lambda=\pm 1$ is the band index, $\theta=\arctan(k_y/k_x)$, $\mathcal{\s}$ is a measure for the width, $\mathbf{r}=(x,y)$, $x,~y$ are the two space dimensions, $\mathbf{k}=(k_x,k_y)$, $k_x,~ k_y$ represent the $x$ and $y$ momenta respectively. $k_x$ is initialized to one, $k_y$ to zero and $\lambda$ to plus one. In the simulations, we consider a rectangular sheet with periodic boundary conditions on a grid of size $L_x \times L_y= 512 \times 128$ or $20nm \times 5nm$, $A_a$, the external potential is set to zero. The norm of the wave-function, $\norm{\Psi}$, i.e. the probability density, $\rho$ is plotted in Fig.~\ref{fig:schematic}(a) for the initial and a later time-step. For the latter, the wave-packet has been reshaped by curvature, in the case of a flat metric it will only spread, see Fig.~\ref{fig:squezing}(a). The kinematics of relativistic wave-packets for Minkowski space-time is explained in Ref.~\cite{dirac_wavepacket}. 

Defining $\delta g^{ij}(x,y) =C_0 \exp((x/\sigma' )^2+(y/\sigma' )^2)$ where $C_0$ denotes the amplitude and $\sigma'=0.2L_{y}$, the metric is:
\begin{align}
\label{eq:metric}
	g_{ij} = 
	\begin{pmatrix}
		1+\delta g^{11}(x,y) & \delta g^{12}(x,y)   \\
		\delta g^{21}(x,y)  & 1+\delta g^{22}(x,y)   
	\end{pmatrix}.
\end{align}
Here $C_0 < 0$, which constitutes an attractive spatial curvature, resulting in a squeezing effect on the wave-packet as seen in Fig.~\ref{fig:squezing}. In the case of $C_0 > 0$ the wave-packet would be repelled from the central region splitting it up. For numerical stability we keep $|C_0|<0.1 L_y$ , additionally small out of plane deformations are more easily physically realizable. 

Experimentally, optical forging can be used to construct graphene into the free-standing three-dimensional shape \cite{forging_graphene}. The method exploits local strain induction due to irradiation with laser pulses under inert atmosphere and has been shown to form up to $20nm$ high custom made structures. The metric tensor can be computed from the discreet mapping (or chart) $h^\alpha(x,y) $ relating the positions of the atoms from the three dimensional flat space (laboratory frame with Minkowski-metric) to the curved space by:
\begin{equation}
    g_{ij}= \frac{\del h^\alpha(x,y)}{\del x^i}\frac{\del h^\beta(x,y)}{\del x^j}\eta_{\alpha \beta},
\end{equation}
as explained in Ref.~\cite{Giordanelli2018}. The positions of the atoms and consequently $h^\alpha$ can be determined by scanning tunneling microscopy and atomic force microscopy. 

The amount of spatial curvature is best described by the Ricci scalar $R$, shown in Fig.~\ref{fig:schematic}(b). 
$R$ represents the contraction of the Riemann curvature tensor $R=g^{ij}R^k_{ikj}$, see Appendix~\ref{app:riemannian}. 
The space averaged Ricci scalar $\langle R \rangle$ is calculated from 
\begin{equation}
\langle R \rangle= \bigg( \int\limits^{x,y} R(x,y) \sqrt{g}dx dy\bigg) / \int\limits^{x,y} \sqrt{g}dx dy,  
\end{equation}
where an explicit expression can be found in the Appendix~\ref{app:ricciscalar}.

The wave-packet undergoes some spreading perpendicular to the motion as expected qualitatively from the Dirac equation \cite{dirac_wavepacket}, shown in Fig.~\ref{fig:squezing}(a). The effect of curvature is to squeeze the wave-packet along the zero momentum direction as shown in Fig.~\ref{fig:squezing}(b) and (c). This is not a geometrical artifact as the squeezed shape is retained, with some spreading, even after the wave-packet exits the curved region.

\begin{figure}
\includegraphics[width=\columnwidth]{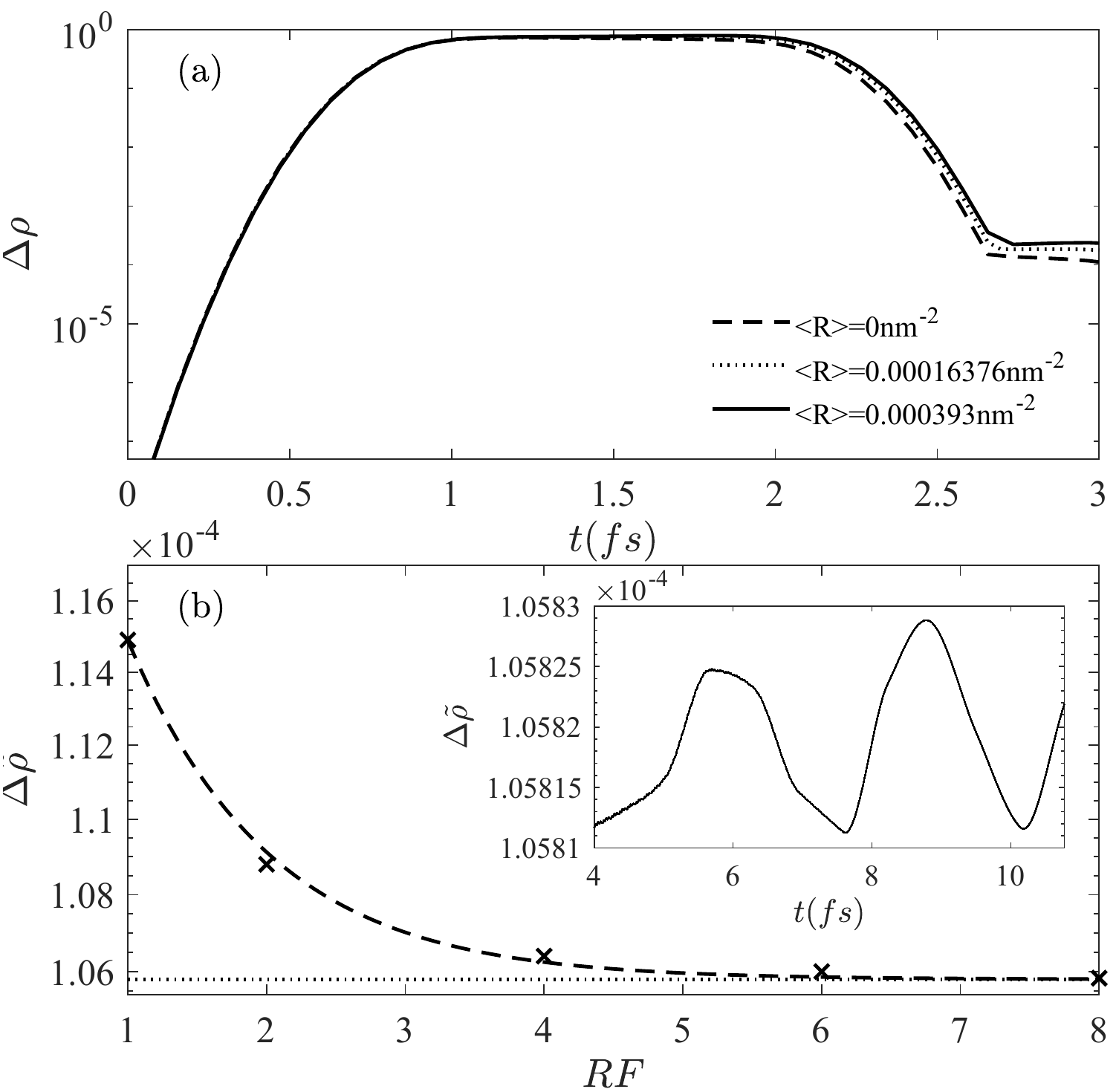}
\caption{\label{fig:resstudy} \textbf{(a)} Time-evolution of  total relative change  of the probability density in region $II$ for different average curvatures $\langle R \rangle$, in semi-logarithmic scale. \textbf{(b)}  Density change in region $II$ for $t=3.0fs$ and $\langle R \rangle= 3.9 \times 10^{-3}nm^{-2}$ plotted against different resolutions as $x$ data points. The residual density is exponentially converging with resolution to an asymptotic value shown as the dotted line. The curve $2.5 \times 10^{-5}\exp(-1.006RF) + 1.058 \times 10^{-4}$ is fitted to the data as the dashed line. The inset shows the probability density within region $II$ for $\langle R \rangle=3.9 \times 10^{-3}nm^{-2}$ and $RF=8$ after a long time.}
\end{figure}

The relative change in normalized probability density,
\begin{equation}
\Delta\rho(t)= \sum\limits^{x,y ~ \in ~ II} \big[ \rho(x,y,t) - \rho(x,y,t=0) \big],
\end{equation}
within the central region $II$, as indicated in Fig.~\ref{fig:schematic}(a), is measured for all time-steps and plotted in logarithmic scale in Fig.~\ref{fig:resstudy}(a). The main bulk of the wave-packet exits the region $II$ at around $t=2.6fs$. Comparing the flat with the curved cases, one sees that some density is left within region $II$. 

In order to exclude the possibility of a numerical artifact, the residual density relative to the flat case at $t=3fs$,
\begin{equation}
\Delta \tilde{\rho}= \Delta\rho(t=3fs) - \Delta\rho_{\langle R \rangle=0}(t=3fs), 
\end{equation}
is plotted in logarithmic scale for $\langle R \rangle =3.9 \times 10^{-3}nm^{-2}$ against the \textit{resolution  factor (RF)} $=$ (number of computational cells)$/$(smallest number of computational cells), for the same physical scenario. From Fig.~\ref{fig:resstudy}(b), by fitting the exponential $2.5 \times 10^{-5}
\exp(-1.006RF) + 1.058 \times 10^{-4}$, we conclude that the residual density is exponentially converging with resolution to an asymptotic value. Therefore the density confining effect is a real solution and not a numerical artifact. 

\begin{figure}
\includegraphics[width=\columnwidth]{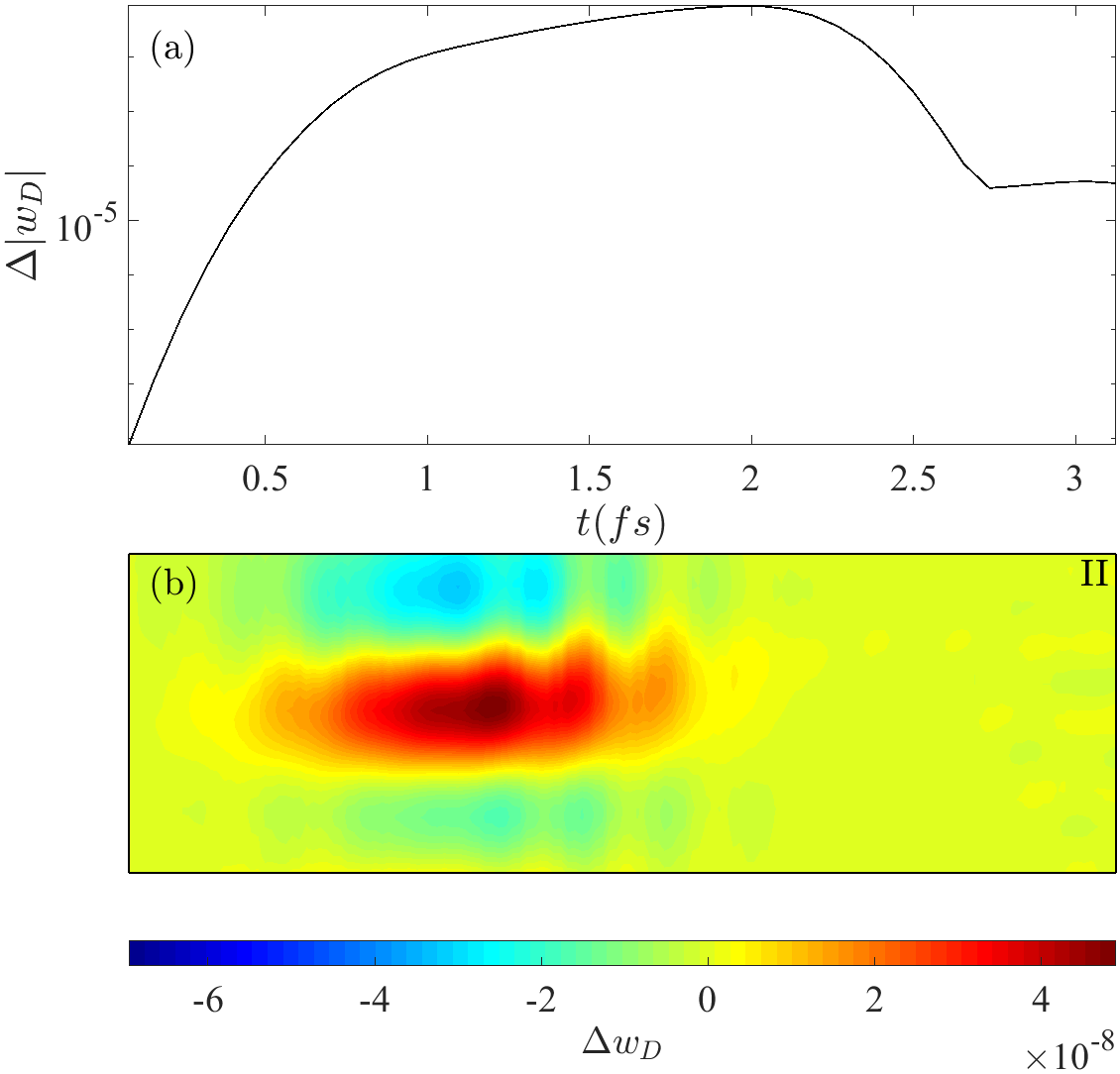}
\caption{\label{fig:angmomemntum}\textbf{(a)} Time-evolution of total vorticity change, a measure of angular momentum, in region $II$ for $\langle R\rangle= 3.9 \times 10^{-3}nm^{-2}$, in semi-logarithmic scale. \textbf{(b)} Total vorticity within region $II$ for $t=2.9fs$ and $\langle R \rangle= 3.9 \times 10^{-3}nm^{-2}$.}
\end{figure}

In Ref.~\cite{relwavepacket}, the vorticity or angular momentum of a relativistic wave-packet is defined as $w_D=\nabla \times J^\mu=\nabla \times \overline{\Psi}\gamma^{\mu} \Psi$. Therefore, as a measure of the angular momentum the total vorticity of the wave-packet relative to the flat scenario
\begin{multline}
\Delta |w_D(t)|= 
\\
\sum\limits^{x,y ~ \in ~ II} \big[ |w_D|(x,y,t) - |w_D|(x,y,t=0) \big] -
\\
\sum\limits^{x,y ~ \in ~ II} \big[ |w_D|(x,y,t) - |w_D|(x,y,t=0) \big]_{\langle R \rangle=0} , 
\end{multline}
within the region II is plotted against time in Fig.~\ref{fig:angmomemntum}(a). Similarly to the probability density, some residual angular momentum remains confined even after the wave-packet exits the region. The confined $\Delta w_D(t=2.9fs)$ is plotted in Fig.~\ref{fig:angmomemntum}(b).

\begin{figure}
\includegraphics[width=\columnwidth]{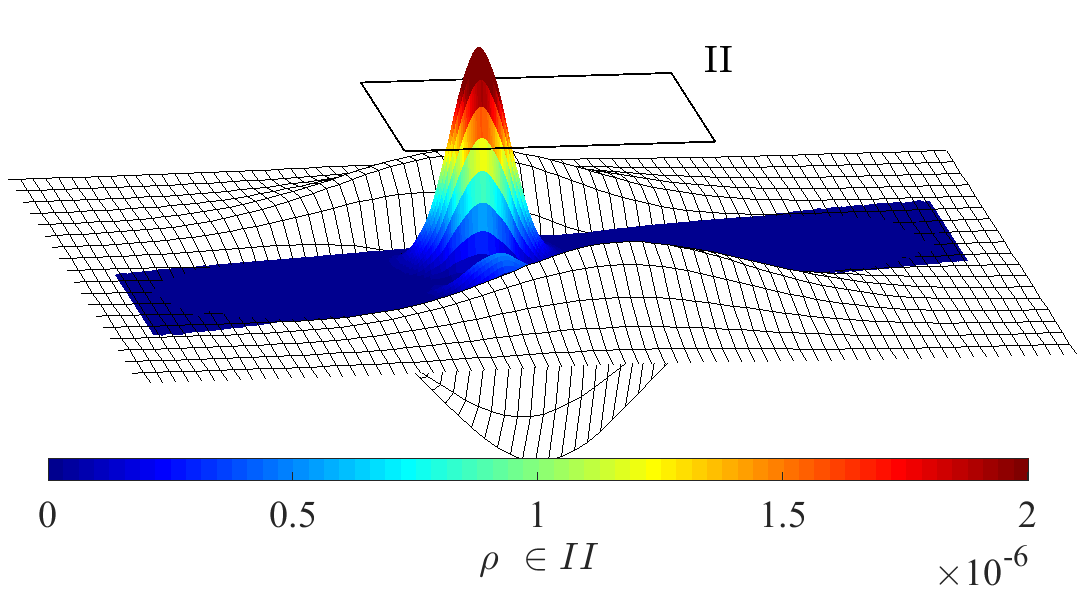}
\caption{\label{fig:randrho}  Normalized confined density  within region $II$, indicated by the top rectangle,  for $t=2.9fs$ and $\langle R \rangle= 3.9 \times 10^{-3}nm^{-2}$.  The curved wire-frame represents the Ricci scalar of the domain. }
\end{figure}

The density confined in the region $II$ at $t=2.9fs$ is  plotted over a wire-frame visualization of the Ricci scalar in Fig.~\ref{fig:randrho}. The density retains a squeezed Gaussian shape with a higher probability in the center, similar to the forward moving wave-packet. The  trapped Dirac fermion density can be understood as the probability of confining a Dirac fermion within the curved space region. Therefore, a Gaussian peak is the natural, expected shape of the confined density. For relatively small curvatures, as investigated here, this probability is about $0.1\%$. Alternatively the wave-packet can be apprehended as a collection of Dirac fermions, which can break apart and some remain within the 'curved space trap'. 

\begin{figure}
\includegraphics[width=\columnwidth]{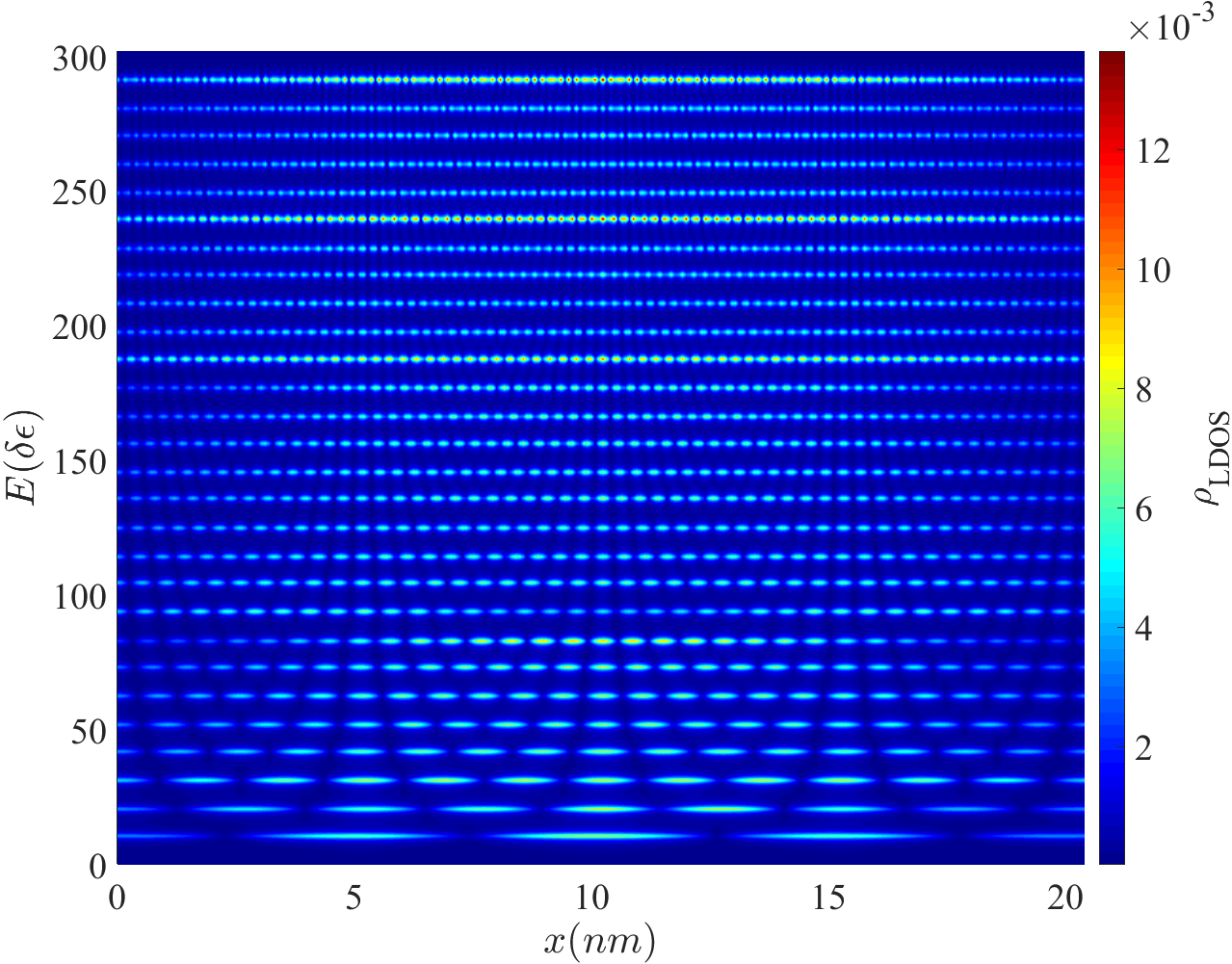}
\caption{\label{fig:LDOS}\textbf{} Local density of states plotted for energy along the x dimension for $y=0.5L_y$.}
\end{figure}

The local density of states (LDOS) is calculated from the energy spectrum $E_n$ of the system and its normalized eigenfunctions $\phi_n(x)$ according to the following relation:
\begin{equation}
    \rho_{LDOS}(x,E)=\frac{1}{\pi}\sum_n |\phi_n(x)|^2 \Im \frac{1}{E-E_n-i\delta \epsilon}.
\end{equation}
$\delta \epsilon \approx 0.02eV$ is the approximate broadening of the energy spectrum peaks and is  expected to be present in the material. The result is plotted in Fig.~\ref{fig:LDOS}, where the discrete energy levels of the system are clearly visible.

The stability of the confined density is investigated by simulating a longer time.
This result is shown in the inset of Fig.~\ref{fig:resstudy}(b). The confined $\Delta \tilde{\rho}$ remains constant for up to $1.2\times 10^3$ computational time-steps and, within numerical errors, showing no indications of depletion. The oscillations of $\mathcal{O}(10^{-5})$ relative to $\Delta \tilde{\rho}$, are caused by minor boundary effects since their period of oscillation is dependent on domain size. In real units, using graphene as an example, for deformation of $20nm$ and Fermi velocity $v_f \approx 1\times 10^6m/s$ \cite{fermi_velocity}, one computational time-step corresponds to $0.01fs$ of physical time. The total simulation time is then equivalent to $12fs$.

\begin{figure}
\includegraphics[width=\columnwidth]{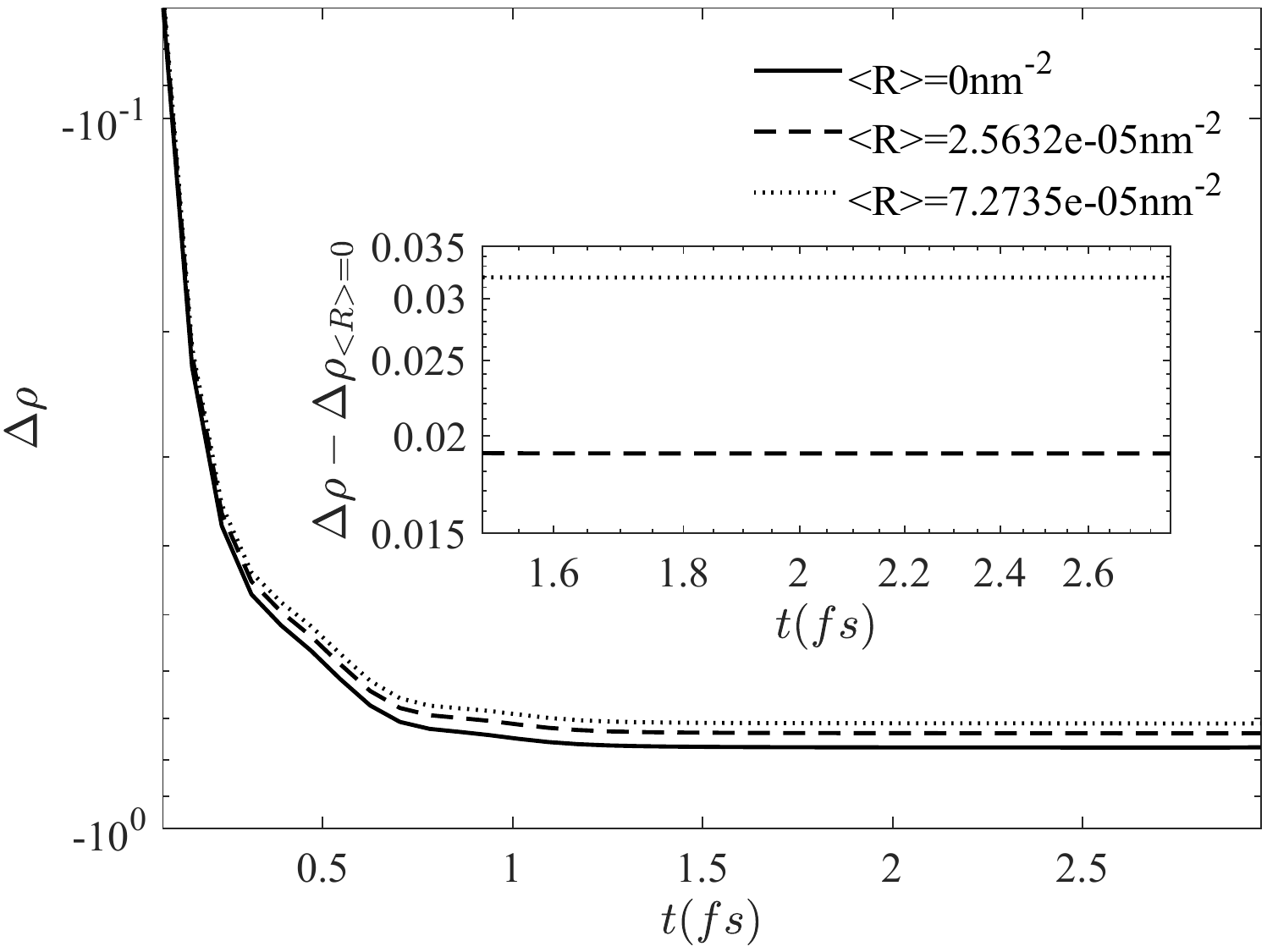}
\caption{\label{fig:zeromomentum}\textbf{} Wave-packet at rest, time-evolution of total relative change  of the probability density in region $II$ for different average curvatures $\langle R \rangle$, in semi-logarithmic scale. The inset shows the residual of the difference between the curved and flat geometries in logarithmic scale.}
\end{figure}

For completeness, 
the same wave-packet is initialized at rest ($k_x=k_y=0$) at the center of region $II$. The time-evolution of the relative change in normalized probability density within region $II$ is plotted in Fig.~\ref{fig:zeromomentum}. The wave-packet spreads outwards but similarly to the previous scenario ($k_x \approx 1$) there is some residual density in region $II$ for the curved relative the to the flat space. As seen in the inset of Fig.~\ref{fig:zeromomentum}, the difference between the curved and flat space remains constant indicating confined charge density in the region.

The present  model describes perfect Dirac fermions and a pure mono-layer graphene sheet for low energy levels close to the Dirac point. In experimental reality the result should be stable to disorder significantly smaller to the curve space trap. For example, ripples of the order of $0.1nm$ would not affect the trapping.   Equivalently, any impurities and/or dislocations would only become significant if they alter basic properties of graphene such as lattice periodicity affecting dramatically the phase space representation.

\begin{figure}
\includegraphics[width=\columnwidth, height=\columnwidth]{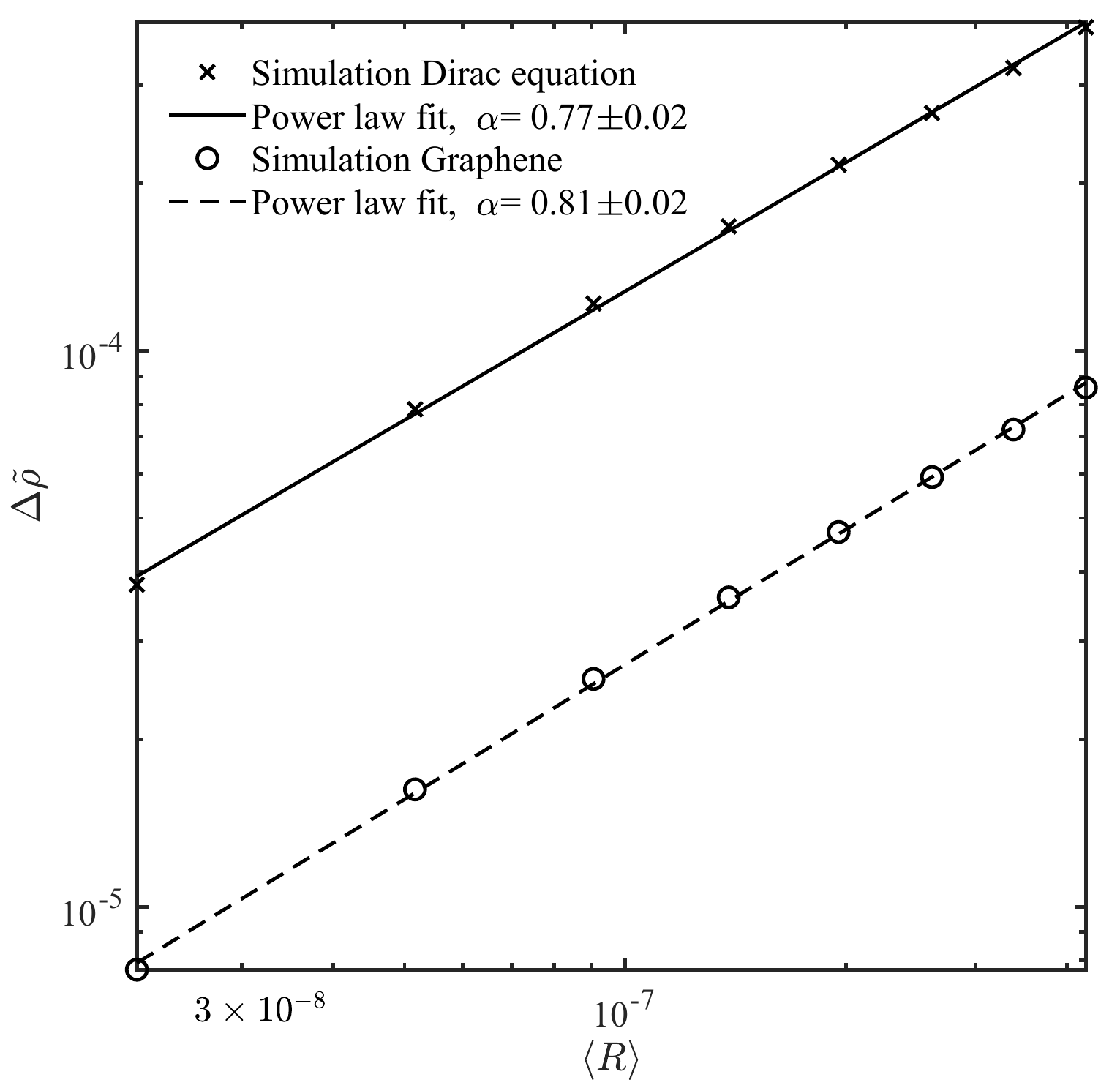}
\caption{\label{fig:maincurve} Normalized confined density in region $II$ plotted against average curvature, in logarithmic scale, for the pure Dirac and the effective graphene Hamiltonians. Power law fit with its fitting error, where $\alpha$ denotes the exponent.}
\end{figure}

The dependence of confinement on curvature is plotted in Fig.~\ref{fig:maincurve} and fitted with a power law. Specifically, the residual density relative to the flat case at $t=3.0fs$ is plotted against total average curvature $\langle R \rangle$.
From Fig.~\ref{fig:maincurve}, an explicit expression for the probability of confinement, $\mathcal{P}_c$ can be deduced:  
\begin{equation}
\label{eq:probc}
\mathcal{P}_c=\frac{\Delta \norm{\Psi}}{\langle \norm{\Psi} \rangle} \propto \frac{\langle R \rangle^\alpha} {\langle \norm{\Psi} \rangle},
\end{equation}
where $\Delta \norm{\Psi}$ denotes the change in probability density and $\langle R \rangle$ the space averaged Ricci scalar.
Eq.~(\ref{eq:probc}) is only valid for an attracting and confining curved space manifold, the exponent $\alpha$ is also case specific,  where we find in the current scenario $\alpha=0.77$  and $\alpha=0.81$  for the pure Dirac Eq.~(\ref{eq:dirac_hamiltonian}) and the graphene effective Eq.~(\ref{eq:graphene_hamiltonian}) Hamiltonians respectively. The discrepancy between the two cases is expected due to the differences in the models, pure Dirac particles are less probable to be confined relative to electrons on graphene.

\section{Summary and Outlook}

We presented a study on transmittance and confining of Dirac particles in 2-D curved space and graphene sheets showing that curvature can squeeze a traveling wave-packet. 

Furthermore, we have shown that it is possible to confine a portion of a traveling wave-packet within a curved space region on a 2D manifold. We propose Eq.~(\ref{eq:probc}) for describing the probability of confinement. In principle, this effect could be experimentally verified by forging graphene into a curved shape \cite{forging_graphene} to reproduce the metric in Eq.~(\ref{eq:metric}). 

Building on these results, other geometries and even time-dependent metrics can be further investigated to increase confinement probability and lifetime. Locally confined Dirac fermions on graphene might be potentially utilized in advanced electronics applications such as memory modules and quantum computing. Additionally, by experimenting further with possible geometries and their effect on traveling wave-packets and currents, a viable graphene wave-guide might be numerically engineered.

For this study we implemented a curved space Dirac equation solver \cite{JD_thesis,jd_paper} based on the quantum Lattice Boltzmann method \cite{succi_qlbm}. This solver can be further developed to curved space-time, opening up the possibility of numerically solving quantum field theories in curved space time towards combining quantum field theory with general relativity.


\begin{acknowledgments}
The authors are grateful for the financial support of the ETH Zurich under Grant No. 06 11-1, and the European Research Council (ERC) Advanced Grant 319968-FlowCCS.
\end{acknowledgments}

\bibliographystyle{ieeetr}
\bibliography{allcitations.bib}


\appendix
\section{Numerical model: Quantum Lattice Boltzmann \label{sec:QLBM}}
\label{app:method}

The Dirac equation in curved space can be written as  
\begin{equation}
\del_t \Psi + \s^a\del_a \Psi = \mathcal{C} \Psi + \mathcal{F} \psi,
\end{equation}
where the left hand side represents the 'free streaming' step along matrix valued 'velocities' $\s^i$ and the right hand site contains a 'collision term' 
\begin{equation}
\label{eq:collision}
\mathcal{C}=-(i m \g^0 + \s^a e_a^i\G_i), 
\end{equation}
where m is the fermion mass, and a 'forcing term' 
\begin{equation}
\mathcal{F}=-\s^a(e_a^i-\d_a^i) \del_i.
\end{equation}
As explained in Ref.~\cite{jd_paper} to avoid interpolation the partial derivative is distributed between the streaming part and the forcing term resulting in a lattice compatible streaming operator of the form $\del_t + v^a\del_a$. 
In order to obtain a diagonal streaming operator the complex $\s$-matrices have to be diagonalized first, which also yields a diagonal velocity matrix with eigenvalues $v^a=\pm 1$. The digitalization is achieved by:

\begin{align*}
	X_c^\dagger \,\a^c\, X_c  
	= \begin{pmatrix}
			 1 & 0 & 0 & 0 \\
			 0 & 1 & 0 & 0 \\
			 0 & 0 & -1 & 0 \\
			 0 & 0 & 0 & -1
		\end{pmatrix}
	= \g^0 \qquad \text{for } c=0,1,2,
\end{align*}
with unitary transformation matrices $X_1, X_2$ given by
\begin{align*}
	\XX_1 &= \T\frac{1}{\sqrt 2} 
		\begin{pmatrix}
			 1 & 0 & -1 & 0 \\
			 0 & 1 & 0 & -1 \\
			 0 & 1 & 0 & 1 \\
			 1 & 0 & 1 & 0
		\end{pmatrix},\\
	\XX_2 &= \T\frac{1}{\sqrt 2} 
		\begin{pmatrix}
			 0 & i & 0 & 1 \\
			 -i & 0 & i & 0 \\
			 -1 & 0 & -1 & 0 \\
			 0 & -1 & 0 & -i
		\end{pmatrix}.
\end{align*}

 The streaming and collision operations are performed in successive steps using operator splitting as simultaneous diagonalization of three $\s$ matrices is not possible, please refer to Ref.~\cite{jd_paper} for the operator splitting procedure. Here, the collision operator is expanded in a unitary way to conserve the norm but since the streaming and forcing terms contain derivative operators a unitary expansion is not possible. Therefore, a simple $2^{nd}$-order expansion is performed limiting the probability norm to $\Delta t^2$ accuracy. 

The manifold itself is described by a chart h defined in linear space (see Appendix.~\ref{app:riemannian}), which is discretized on a regular rectangular lattice. The curved space quantum Lattice Boltzmann method evolves the spinor $\Psi = (\Psi^+, \Psi^-) = (\Psi_1^+,\Psi_2^+,\Psi_1^-,\Psi_2^-)$ from $t$ to $t+\delta t$. Once the operators are split, the following algorithm is performed consecutively for each lattice direction $n_a$, where $n_1=(1,0)$, $n_2=(0,1)$ and $a=1,2$. 

\begin{enumerate}
\item \textbf{Rotation:} The spinor is rotated by $X_a$ 
\begin{equation}
\tilde{\Psi}_a(x,t)=X^\dagger_a \Psi(x,t).
\end{equation}

\item \textbf{Collisions and curvature:} The collision and force operators are applied on the rotated spinor,
\begin{equation*}
\tilde{\Psi}_a^*(x,t)=\big( \Delta t \tilde{\mathcal{F}}_a + ( \mathbb{1} - \frac{\Delta t}{2} \tilde{\mathcal{C}}_a)^{-1}  ( \mathbb{1} + \frac{\Delta t}{2} \tilde{\mathcal{C}}_a)   \big) \tilde{\Psi}_a(x,t),
\end{equation*}
where $\tilde{\Psi}_a^*(x,t)$ denotes an auxiliary field,
\begin{equation}
\label{eq:collisiontilde}
\tilde{\mathcal{C}}_a=\frac{1}{2}X^\dagger_a \mathcal{C}X_a,
\end{equation}
\begin{equation}
\label{eq:forcingtilde}
\tilde{\mathcal{F}}_a\tilde{\Psi}_a(x,t)=\big(e_a^i-\delta_a^i\big) \Big(\tilde{\Psi}_a(x \mp n_i \Delta t,t)-\tilde{\Psi}_a(x,t) \Big),
\end{equation}
$n_i$ the lattice direction and $\mathcal{C}$ the collision term, Eq.~(\ref{eq:collision}).
The upper sign applies for the spin-up components $(\Psi_1^+,\Psi_2^+)$ and the lower sign for the spin-down components $(\Psi_1^-,\Psi_2^-)$.
\item \textbf{Streaming:} The spinor components are streamed to the closest grid points along the lattice direction $\pm n_a$,
\begin{equation}
\tilde{\Psi}_a(x,t+\frac{\Delta t}{2})=\tilde{\Psi}_a^*(x \mp n_a\Delta t,t).
\end{equation}

\item \textbf{Inverse Rotation:} The spinor is rotated back by $X_a$,
\begin{equation}
\Psi_a(x,t+\frac{\Delta t}{2})=X_a \tilde{\Psi}_a(x,t+\frac{\Delta t}{2}).
\end{equation}

\item Repeat steps 2-4 for the next spatial direction
\end{enumerate}

The simulation for strained graphene is carried out with modified Eqs.~(\ref{eq:collisiontilde},\ref{eq:forcingtilde}) such that 
\begin{equation*}
\tilde{\mathcal{C}}_a \rightarrow \sqrt{g} \tilde{\mathcal{C}}_a, ~ e_a^i\rightarrow \sqrt{g}e_a^i.
\end{equation*}
The additional factor of $\sqrt{g}$ originates from the volume element of the Hamiltonian Eq.~(\ref{eq:hamiltoniangraphene}).

\section{Riemannian geometry}
\label{app:riemannian}

The Latin indices run over the spatial dimensions and Einstein summation convection is used for repeated indices.

A $D$ dimensional curved space is represented by a Riemannian manifold M, which is locally described by a smooth diffeomorphism $\mathbf{h}$, called the chart. The set of tangential vectors attached to each point $\mathbf{y}$ on the manifold is called the  tangent space $T_\mathbf{y} M$. In the fluid model, all the vector quantities are represented as elements of $T_\mathbf{y} M$. The derivatives of the chart $\mathbf{h}$ are used to define the standard basis $(\textbf{e}_1,...,\textbf{e}_D)=\frac{\del\mathbf{h}}{\del x^1},...,\frac{\del \mathbf{h}}{\del x^D}$. 

The metric tensor $g$ can be used to measure the length of a vector or the angle between two vectors. In local coordinates, the components of the metric tensor are given by 
\begin{equation}
g_{ij}(x)= \textbf{e}_i(x)\cdot \textbf{e}_j(x)= \frac{\del \mathbf{h}}{\del x^i} \cdot \frac{\del \mathbf{h}}{\del x^j},
\end{equation}
where $\cdot$ is the standard Euclidean scalar product.

For a given metric tensor, the vector $v=v^i\textbf{e}_i \in T_\mathbf{y} M$ has a norm $||v||_g=\sqrt{v^ig_{ij}v^j}$ and a corresponding dual vector $v^*=v^i\textbf{e}_i \in T^*_\mathbf{y} M$ in the cotangent space, which is spanned by the differential 1-forms $dx^i=g(\textbf{e}_i,\cdot)$. The coefficients $v_i$ of the dual vector are typically denoted by a lower index and are related to the upper-index coefficients $v^i$ by contraction with the metric tensor $v_i = g_{ij}v^j$ or equivalently, $ v^i=g^{ij}v_j$, where $g^{ij}$ denotes the inverse of the metric tensor. The upper-index coefficients $v^i$ of a vector $v$ are typically called  \textit{contravariant components}, whereas the lower-index coefficients $v_i$ of the dual vectors $v^*$ are known as the \textit{covariant components}.

A necessary feature for the description of objects moving on the manifold is parallel transport of vectors along the manifold.  The tangent space is equipped with a covariant derivative $\nabla$ (Levi-Civita connection), which connects the tangent spaces at different points on the manifold and thus allows to transport a tangent vector  from one tangent space to the other along a given curve  $\gamma(t)$. The covariant derivative can be viewed  as the orthogonal projection of the Euclidean derivative $\del$ onto the tangent space, such that the tangency of the vectors is preserved during the transport. In local coordinates, the covariant derivative is fully characterized by its connection coefficients $\Gamma^i_{jk}$  (Christoffel symbols), which are defined by the action of the covariant derivative on the basis vector, $\nabla_j \textbf{e}_k= \Gamma^i_{jk}$. In the standard basis, $\textbf{e}_i = \frac{\del \mathbf{h}}{\del x^i}$, the Christoffel symbols are related to the metric by
\begin{equation}
\Gamma^i_{jk}=\frac{1}{2}g^{ij}(\del_j g_{kl} + \del_k g_{jl} - \del_l g_{jk}).
\end{equation} 
Acting on a general vector $v=v^i \mathbf{e}_i,$ the covariant derivative becomes:
 \begin{equation}
\nabla_k v =(\del_k v^i + \Gamma^i_{kj}v^j)\mathbf{e}_i,
\end{equation}
where the product rule has been applied, using that the covariant derivative acts as a normal derivative on the scalar functions 
$v^i$. Extending to tensors of higher rank, for example the second order tensors $T= T^{ij} $, 
\begin{equation}
\nabla_k T=( \del_kT^{ij}+ \G ^i _{kl} T^{lj} + \G^j_{kl} T^{il})\mathbf{e}_i \otimes \mathbf{e}_j
\end{equation}
in this work the basis vectors $\mathbf{e}_i $ are generally dropped. Compatibility of the covariant derivative with the metric tensor implies that $\nabla_k g^{ij}=\nabla_k g_{ij} =0$. This property allows us to commute the covariant derivative with the metric tensor for the raising or lowering of tensor indices in derivative expressions.

The motion of the particle can be described by the curve $ \gamma(t)$, which parametrizes the position of the particle at time $t$. The geodesic equation,  $ \nabla_{\dot{\gamma}} \dot{\gamma} =0 $, in local coordinates $\gamma(t)=\gamma^i(t)\mathbf{e}_i$ is defined by
\begin{equation}
\label{eq:geodesic}
\ddot{\gamma}^i + \Gamma_{jk}^i \dot{\gamma^j} \dot{\gamma^k} = 0. 
\end{equation}
The geodesic equation can be interpreted as the generalization of Newtons law of inertia to curved space. The solutions of Eq.~(\ref{eq:geodesic}) represent lines of constant kinetic energy on the manifold, i.e. the geodesics. 
	The Riemann curvature tensor $R$ can be used to measure curvature, or more precisely, it measures curvature-induced change of a tangent vector $v$ when transported along a closed loop.
 \begin{equation}
R(\textbf{e}_i,\textbf{e}_j)v=\nabla_i \nabla_j v-\nabla_j \nabla_i v.  
\end{equation}   
In a local coordinate basis $ { \textbf{e}_i } $, the coefficients of the Riemann curvature tensor are given by
\begin{multline}
R^l_{ijk}= g(R(\textbf{e}_i,\textbf{e}_j)\textbf{e}_k,\textbf{e}_l) =
\\
\del_j \Gamma^l_{ik} - \del_k \Gamma^l_{ij} + \Gamma^l_{jm} \Gamma^m_{ik} -\Gamma^l_{km} \Gamma^m_{ij}.
\end{multline}
Contraction of $R^i_{jkl}$ to a rank 2 and 1 tensor yields  the Ricci-tensor $R_{ij}=R^k_{ikj}$ and the Ricci-scalar $R=g^{ij}R_{ij}$ respectively, which can also be used to quantify curvature.

The gradient is defined as $\nabla^i f= g^{ij} \del_j f$, the divergence as $\nabla_i v^i= \frac{1}{\sqrt{g}} \del_i (\sqrt{g} v^i) $, and the integration over curved volume as $V=\int_V QdV$, where $dV=\sqrt{g}dx^1...dx^D=:\sqrt{g}d^Dx$ denotes the volume element. $\sqrt{g}$ denotes the square root of the determinant of the metric tensor.   

It should be clarified that in the simulations there is no time curvature and $g_{ij}$ denotes the curved space metric.

\section{Explicit Ricci Scalar }
\label{app:ricciscalar}
The Ricci scalar $R$ as explicitly calculated for the metric Eq.~(\ref{eq:metric}):
\begin{multline*}
R=-(C_0 e^{-\frac{7}{4} \left(x^2+y^2\right)} \bigg[ 2 C_0^4 
   \left(e^{\frac{3}{4} \left(x^2+y^2\right)}-4\right)^2 \times
   \\
   \left(e^{\frac{3}{4} \left(x^2+y^2\right)}+1\right)  (x-y)^2+C_0^3 e^{\frac{1}{4} \left(x^2+y^2\right)} \times
   \\
   \big(e^{3 \left(x^2+y^2\right)} (x-y)^2+e^{\frac{9}{4}
   \left(x^2+y^2\right)} \times
   \\
   \left(x^2-2 x y+y^2-32\right)-2 e^{\frac{3}{2} \left(x^2+y^2\right)} \times
   \\
   \left(3 x^2-6 x y+3 y^2+80\right)+4 e^{\frac{3}{4} \left(x^2+y^2\right)}\times
   \\
   \left(49 x^2-98 x y+49 y^2-32\right)+
   \\
   16(x-y)^2 \big)+C_0^2 e^{\frac{5}{4} \left(x^2+y^2\right)} \times
   \\
   \big(2  e^{\frac{9}{4} \left(x^2+y^2\right)} \left(3 x^2-6 x y+3 y^2-8\right)+
   \\
   32 \left(3 x^2-6 x y+3 y^2-2\right)+16 e^{\frac{3}{4} \left(x^2+y^2\right)} \times
   \\
   \left(10 x^2-20xy+10 y^2-17\right)+e^{\frac{3}{2} \left(x^2+y^2\right)} \times
   \\
   \left(55 x^2-110 x y+55 y^2-128\right) \big)+C_0 e^{\frac{9}{4} \left(x^2+y^2\right)} \times
   \\
   \big(e^{\frac{3}{2} \left(x^2+y^2\right)} \left(5 x^2-10 x y+5 y^2-24\right)+
   \\
   16 \left(5 x^2-10 x y+5 y^2-6\right)+2 e^{\frac{3}{4} \left(x^2+y^2\right)} \times
   \\
   \left(47x^2-94 x y+47 y^2-60\right)\big)+2 e^{4 \left(x^2+y^2\right)} \times
   \\
   \left(x^2-2 x y+y^2-4\right)+32 e^{\frac{13}{4} \left(x^2+y^2\right)} \times
   \\ 
   \left(x^2-2 x y+y^2-1\right))\bigg]/
   \\
   (16 \left(2 C_0+e^{\frac{1}{4}
   \left(x^2+y^2\right)}\right)^2 \left(2 C_0+e^{x^2+y^2}\right)^2).
\end{multline*}
\end{document}